\documentclass[10pt,conference]{IEEEtran}
\usepackage{graphicx,amssymb,amsmath}
\IEEEoverridecommandlockouts                              
\newtheorem{defi}{Definition}
\newtheorem{thm}{Theorem}
\newtheorem{prop}{Proposition}
\newtheorem{corr}{Corollary}

\newcommand{\be}{\begin{equation}}
\newcommand{\ee}{\end{equation}}
\newcommand{\ben}{\begin{equation*}}
\newcommand{\een}{\end{equation*}}

\title{On Evaluating the Rate-Distortion Function of Sources with Feed-Forward and the Capacity of Channels with Feedback}
\author{
\authorblockN{Ramji Venkataramanan and S. Sandeep Pradhan}
\authorblockA{Department of EECS,
University of Michigan,
Ann Arbor, MI 48105 \\
rvenkata@umich.edu, pradhanv@eecs.umich.edu}
\thanks{This work was supported by NSF ITR Grant CCF-0427385 and Grant (CAREER) CCF-0448115.}}

\begin{document}

\maketitle

%

\begin{abstract}
In this work, we study the problem of evaluating the performance
limit of two communication problems that are closely related to each
other- source coding with feed-forward and channel coding with
feedback. The formulas (involving directed information) for the
optimal rate-distortion function with feed-forward and channel
capacity with feedback are multi-letter expressions and cannot be
computed easily in general. In this work, we derive conditions under
which these can be computed for a large class of sources/channels
with memory and distortion/cost measures. Illustrative examples are
also provided.
\end{abstract}
\section{Introduction}
Feedback is widely used in communication systems to help combat the
effect of noisy channels. It is  well-known that feedback does not
increase the capacity of a discrete memoryless channel
\cite{shannon}. However, feedback could increase the capacity of a
channel with memory. Recently, directed information has been used to
elegantly characterize the capacity of channels with feedback
\cite{massey, kramer,sekar1,tati06}.  The source coding counterpart
of channel coding with feedback is source coding with feed-forward.
Channels with feedback have been studied extensively, but the
problem of source coding with feed-forward is recent
\cite{tsa,pradhan07,mw,rdff}.

 Source coding with
feed-forward can be explained in simple terms as follows. In the
usual fixed-rate lossy source coding problem, there is a source $X$
that has to be reconstructed at a decoder with some distortion $D$.
The encoder takes a block of, say, $N$ source samples and maps it to
an index in a codebook. The decoder uses this index to generate the
reconstruction of the $N$ source samples. In source coding with
feed-forward, the encoder works in a similar fashion and sends an
index to the decoder.
The decoder
generates the reconstructions  sequentially: in order to reconstruct
each source sample, the decoder has access to the index as well as
some past source samples. More precisely, let $X_n,\hat{X}_n$ denote
the source and reconstruction samples at time $n$, respectively. If
the source samples are available with a delay $k$ after the index is
sent, to generate $\hat{X}_n$, the decoder has knowledge of the
index plus the source samples until time $n-k$. This problem is
called feed-forward with delay $k$, and it is of interest to study
the rate vs. distortion trade-offs in this setting \cite{tsa,rdff}.

Source coding with feed-forward was considered in the context of
competitive prediction in \cite{tsa}. The problem was motivated and
studied in \cite{pradhan07, mw, rdff} from a communications
perspective, as a variant of source coding with side information.
For instance, we can consider the source to be a field that needs to
compressed and communicated from one node to another in a network.
This field (e.g. a seismic or acoustic field) could propagate
through the medium at a slow rate and become available at the
decoding node as side-information with some delay. Later in this
paper, we will present an example of feed-forward relating to
predicting variations in stock prices.

 The formulas (involving
directed information) for the optimal rate-distortion function with
feed-forward \cite{rdff} and channel capacity with feedback
\cite{sekar1} are multi-letter expressions and cannot be computed
easily in general. In this work, we study the problem of evaluating
the rate-distortion and capacity expressions. We derive conditions
under which these can be computed for a large class of sources
(channels) with memory and distortion (cost) measures. We also
provide illustrative examples. Throughout, we consider source
feed-forward and  channel feedback with arbitrary delay. When the
delay goes to $\infty$, we obtain the case of no
feed-forward/feedback.

\section{Source Coding with Feed-Forward}
\subsection{Problem Formulation}
Consider a general discrete source $X$ with alphabet $\mathcal{X}$,
characterized by a sequence of distributions denoted
$\mathbf{P_X}=\{P_{X^n} \}_{n=1}^{\infty}$. The reconstruction
alphabet is $\widehat{\mathcal{X}}$ and there is  an associated
sequence of distortion measures $d_n:\mathcal{X}^n \times
\mathcal{\widehat{X}}^n \to \mathbb{R}^+$. It is assumed that
$d_n(x^n,\hat{x}^n)$ is normalized with respect to $n$ and is
uniformly bounded in $n$. For example $d_n(x^n,\hat{x}^n)$ may be
the average per-letter distortion, i.e.,
$\frac{1}{n}\sum_{i=1}^{n}d(x_i,\hat{x}_i)$ for some $d: {\mathcal
X} \times \hat{{\mathcal X}} \to \mathbb{R}^+$.
\begin{defi}
An $(N,2^{NR})$ source code with delay $k$ feed-forward  of block
length $N$ and rate $R$ consists of an encoder mapping $e$ and a
sequence of decoder mappings $g_i,  i=1,\ldots,N$, where \ben
\begin{split} e:\mathcal{X}^N & \to \{1,\ldots,2^{NR}\} \\ g_i:
\{1,\ldots,2^{NR}\} \times \mathcal{X}^{i-k} & \to
\widehat{\mathcal{X}},\quad i=1,\ldots,N. \end{split} \een
\end{defi}
 The encoder maps each $N$-length source sequence to an index
in $\{1,\ldots,2^{NR}\}$. The decoder receives the index transmitted
by the encoder, and to reconstruct the $i$th sample ($i>k$), it has
access to the source samples until time ($i-k$) (for $i\leq k$,
$\hat{X}_i$ is produced using the index alone). We want to minimize
$R$ for a given distortion constraint.

\begin{defi}
\emph{(Probability of error criterion)} $R$ is an
$\epsilon$-achievable rate at  distortion $D$ if for all
sufficiently large $N$, there exists an $(N, 2^{NR})$ source
codebook such that
\begin{displaymath}
P_{X^N}\left(x^N: d_N(x^N,\hat{x}^N)>D\right)<\epsilon,
\end{displaymath}
where $\hat{x}^N$ denotes the reconstruction of $x^N$.
$R$ is an achievable rate at probability-$1$ distortion $D$ if it is
$\epsilon$-achievable for every $\epsilon>0$.
\end{defi}

 We now give a brief summary of the rate-distortion results with feed-forward found in \cite{rdff}.
The rate-distortion function with feed-forward (delay $1$) is
characterized by directed information, a quantity defined in
\cite{massey}. The directed information flowing from a random
sequence $\hat{X}^N$ to a random sequence ${X}^N$ is defined as \be
I(\hat{X}^N \to X^N)=\sum_{n=1}^N I(\hat{X}^n ; X_n|{X}^{n-1}).\ee
When the feed-forward delay is $k$, the rate-distortion function is
characterized by the $k-$delay version of the directed information:
\be \label{eq:kdiri} I_k(\hat{X}^N \to X^N)=\sum_{n=1}^N
I(\hat{X}^{n+k-1} ; X_n|{X}^{n-1}).\ee
 When we do not make any assumption on
the nature of the joint process $\{\mathbf{X},\mathbf{\hat{X}}\}$,
we need to use the information spectrum \cite{hv} version of
\eqref{eq:kdiri}. In particular, we will need the
quantity\footnote{The $\limsup_{in prob}$ of a random sequence $A_n$
is defined as the smallest number $\alpha$ such that $\lim_{n \to
\infty} P(A_n > \alpha)=0$ and is denoted $\overline{A}$.}
\be \label{eq:kdirgen} \overline{I}_k(\hat{X} \to X) \triangleq
\limsup_{in prob} \frac{1}{n}\log
\frac{P_{X^n,\hat{X}^n}}{\vec{P}^k_{\hat{X}^n|X^n} \cdot P_{X^n}},
 \ee where \ben
\vec{P}^k_{\hat{X}^n|X^n}=\prod_{i=1}^n
P_{\hat{X}_i|\hat{X}^{i-1},X^{i-k}}.\een
It should be noted that  \eqref{eq:kdiri} and \eqref{eq:kdirgen} are
the same when the joint process $\{\mathbf{X},\mathbf{\hat{X}}\}$ is
stationary and ergodic.
\begin{thm} \emph{\cite{rdff}} \label{thm:dir}
For an arbitrary source $X$ characterized by a distribution
$\bf{P_X}$, the rate-distortion function with feed-forward- the
infimum of all achievable rates at
 distortion $D$- is given by
\begin{equation}\label{eq:dir1}
R_{ff}(D)=\inf_{\mathbf{P_{\hat{X}|X}}:\rho(\mathbf{P_{\hat{X}|X}})
\leq D} \overline{I}_k(\hat{X} \to X),
\end{equation}
where
\small{
\begin{equation}
\begin{split}
&\rho(\mathbf{P_{\hat{X}|X}}) \triangleq \limsup_{in prob}
d_n(x^n,\hat{x}^n)\\
&=\inf \left\{ h: \lim_{n \to \infty} P_{X^n,\hat{X}^n}
\left((x^n,\hat{x}^n):d_n(x^n,\hat{x}^n)>h \right)=0\right\}.
\end{split}
\end{equation}}
\end{thm}

\subsection{Evaluating the Rate-Distortion Function with Feed-forward}
The rate-distortion formula in Theorem \ref{thm:dir} is an
optimization of a multi-letter expression:  \ben
\overline{I}_k(\hat{X} \to X) \triangleq \limsup_{in prob}
\frac{1}{n}\log \frac{P_{X^n,\hat{X}^n}}{\vec{P}^k_{\hat{X}^n|X^n}
\cdot P_{X^n}}, \een
This is an optimization
over an infinite dimensional space of conditional distributions
$\mathbf{P_{\hat{X}|X}}$. Since this is a potentially difficult
optimization, we turn the problem on its head and pose the following
question:

\emph{Given a source $X$ with distribution $\mathbf{P_X}$ and a
conditional distribution $\mathbf{P_{\hat{X}|X}}$, for what sequence
of distortion measures does  $\mathbf{P_{\hat{X}|X}}$ achieve the
infimum in the rate-distortion formula ?}

A similar approach is used in \cite{ck} (Problem 2 and 3, p. 147) to
find optimizing distributions for discrete memoryless channels and
sources without feedback/feed-forward. It is also used in
\cite{gast} to study the optimality of transmitting uncoded source
data over channels and in \cite{duality} to study the duality
between source and channel coding.

 Given a source ${X}$, suppose we have a hunch
about the structure of the optimal conditional distribution. The
following theorem (proof omitted) provides the distortion measures
for which our hunch is correct.

\begin{thm}
\label{thm:scff} Suppose we are given a stationary, ergodic source
$X$ characterized by $\mathbf{P_X}=\{P_{X^n}\}_{n=1}^{\infty}$ with
feed-forward delay $k$. Let
$\mathbf{P_{\hat{X}|X}}=\{P_{X^n|X^n}\}_{n=1}^{\infty}$ be a
conditional distribution such that the joint distribution is
stationary and ergodic. Then $\mathbf{P_{\hat{X}|X}}$ achieves the
rate-distortion function if for all sufficiently large $n$, the
distortion measure satisfies
\be \label{eq:dist_cond} d_n(x^n,\hat{x}^n)=-c \cdot \frac{1}{n}\log
\frac{P_{X^n,\hat{X}^n}(x^n,\hat{x}^n)}{\vec{P}^k_{\hat{X}^n|X^n}(\hat{x}^n|x^n)}+d_0(x^n),
\ee where {\small{
$\vec{P}^k_{\hat{X}^n|X^n}(\hat{x}^n|x^n)= \prod_{i=1}^n
P_{\hat{X}_i|X^{i-k},\hat{X}^{i-1}}(\hat{x}_i|x^{i-k},\hat{x}^{i-1}),
$}} $c$ is any positive number and $d_0(.)$ is an arbitrary
function. The distortion constraint in this case is equal to
$\limsup_{n\to \infty}d_n(x^n,\hat{x}^n)$.
\end{thm}
%

We have considered a conditional distribution
$\mathbf{P_{\hat{X}|X}}$ such that $\mathbf{P_{X}P_{\hat{X}|X}}=
\{P_{X^n}P_{X^n|X^n}\}_{n=1}^\infty$ is stationary, ergodic.
Nevertheless, the theorem gives the condition for optimality of
$\mathbf{P_{\hat{X}|X}}$ among \emph{all} conditional distributions,
not just the ones that make the joint distribution stationary and
ergodic.

\subsection{Markov Sources with Feed-forward}
 A stationary, ergodic $m$th order Markov source ${X}$
is characterized by a distribution
$\mathbf{P_X}=\left\{P_{X^n}\right\}_{n=1}^{\infty}$
 where
 \be \label{eq:marks} P_{X^n}=\prod_{i=1}^n P_{X_i|X^{i-1}_{i-m}}, \quad \forall n. \ee
Let the source have feed-forward with delay $k$. We first ask:
\emph{When is the optimal joint distribution also $m$th order Markov
in the following sense}:
 \be \label{eq:markjoint}
 P_{X^n, \hat{X}^n}= \prod_{i=1}^n P_{X_i,\hat{X}_i|X^{i-1}_{i-m}},
\quad \forall n.
 \ee
 In other words, when does the optimizing conditional distribution
 have the form
 \be \label{eq:markcd}
 P_{\hat{X}^n|X^n}=\prod_{i=1}^n P_{\hat{X}_i|X^{i}_{i-m}}, \quad
 \forall n.
 \ee
The answer, provided by Theorem \ref{thm:scff}, is stated below. We
drop the subscripts on the probabilities to keep the notation clean.

\begin{corr}
\label{corr:a} For an $m$th order Markov source (described in
\eqref{eq:marks}) with feed-forward delay $k$, an $m$th order
conditional distribution (described in \eqref{eq:markcd}) achieves
the optimum in the rate-distortion function for a sequence of
distortion measures $\{d_n\}$ given by {\small{ \be
\label{eq:distmark} d_n(x^n,\hat{x}^n)=-c \cdot \frac{1}{n}
\sum_{i=1}^n \log
\frac{P(x_i,\hat{x}_i|x^{i-1}_{i-m})}{P(\hat{x}_i|\hat{x}_{i-k+1}^{i-1},x_{i-k+1-m}^{i-k})}+d_0(x^n),
\ee}}where $c$ is any positive number and $d_0(.)$ is an arbitrary
function.
\end{corr}
\proof The proof involves substituting \eqref{eq:marks} and
\eqref{eq:markcd} in \eqref{eq:dist_cond} and performing a few
manipulations.
%
%

%
%
\section{Examples}

\subsection{Stock-market example}
\begin{figure} [t]
\begin{center}
\includegraphics[height=1.2in,width=3.4in]{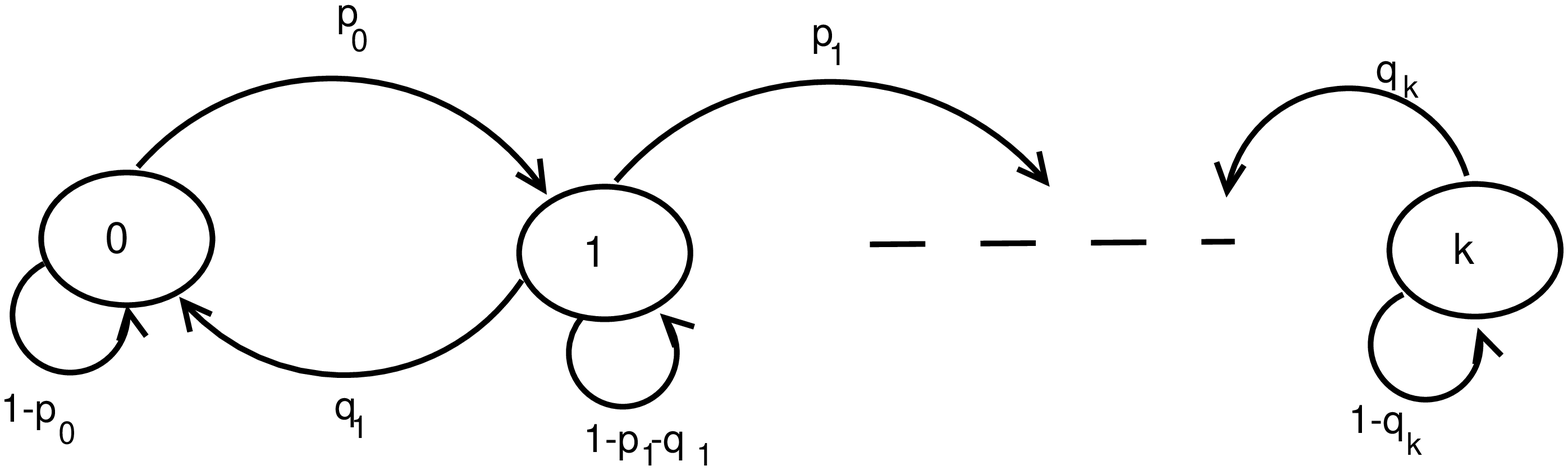}
\end{center}
\caption{Markov chain representing the stock value}
\label{fig:markov}
\end{figure}
Suppose that we wish to observe the behavior of a particular stock
in the stock market over an $N-$day period. Assume that the value of
the stock can take $k+1$ different values and is modeled as a
$k+1$-state Markov chain, as shown in Fig. \ref{fig:markov}. If on a
particular day, the stock is in state $i, \: 1\leq i<k$, then on the
next day, one of the following can happen.
\begin{itemize}
  \item The value increases to state $i+1$ with probability $p_i$.
  \item The value drops to state $i-1$ with probability $q_i$.
  \item The value remains the same with probability $1-p_i-q_i$.
\end{itemize}
When the stock-value is in state $0$, the value cannot decrease.
Similarly, when in state $k$, the value cannot increase.  Suppose an
investor invests in this stock over an $N-$day period and desires to
be forewarned whenever the value drops. Assume that there is an
insider (with some a priori information about the behavior of the
stock over the $N$ days) who can send information to the investor at
a finite rate.

The value of the stock is modeled as a Markov source
$\mathbf{X}=\{X_n\}$. The decision $\hat{X}_n$ of the investor is
binary: $\hat{X}_n=1$ indicates that the price is going to drop from
day $n-1$ to $n$, $\hat{X}_n=0$ means otherwise. Before day $n$, the
investor knows all the previous values of the stock $X^{n-1}$ and
has to make the decision $\hat{X}_n$. Thus feed-forward is
automatically built into the problem.

\begin{table}[t]
\caption{Distortion $e\left(\hat{x}_i,x_{i-1}=j,x_{i}\right)$}
 \label{tab:dist1}
\begin{center}
\begin{tabular}{|c|c|c|c|}
\multicolumn{4}{c}{$ \qquad(x_{i-1},x_i)$}\\
\hline
&$j,j+1$&$j,j$&$j,j-1$ \\
\hline
$\hat{x}_i=0$ &0&0&1 \\
$\hat{x}_i=1$ &1&1&0\\
\hline
\end{tabular}
\end{center}
\end{table}
The investor makes an error either when she fails to predict a drop
or when she falsely predicts a drop. The distortion is modeled using
a Hamming distortion criterion as follows. \be \label{eq:bindist1}
d_n(x^n,\hat{x}^n)= \frac{1}{n} \sum_{i=1}^n
e(\hat{x}_i,x_{i-1},x_i),\ee where $e(.,.,.)$ is the
\emph{per-letter}  distortion given Table \ref{tab:dist1}. The
minimum amount of information (in bits/sample) the insider needs to
convey to the investor so that she can predict drops in value with
distortion $D$ is denoted $R_{ff}(D)$.

\begin{table*}[t]
\caption{The distribution $P\left(X_{i}|x_{i-1},\hat{x}_i\right)$}
\begin{center}
\begin{tabular}{|c|c|c|c|c|c|c|c|c|}
\multicolumn{9}{c}{$ \qquad (x_{i-1},\hat{x}_i)$}\\
\hline
&$0,0$&$0,1$&$\cdots$&$j,0$&$j,1$&$\cdots$&$k,0$&$k,1$ \\
\hline
${x}_i=0$ & $1-p$ & $-$ & $\cdots$ &$-$ &$-$ &$-$ &$-$ &$-$ \\
${x}_i=1$ & $p$ & $-$ & $\cdots$ &$-$ &$-$ &$-$ &$-$ &$-$ \\
${x}_i=\vdots$ &$-$   &$-$  &$\ddots$& $-$  & $-$ &$-$   &$-$  &$-$ \\
${x}_i=j-1$ &$-$   &$-$  &$-$&$\epsilon$ & $1-\epsilon$ &$-$ &$-$ &$-$ \\
${x}_i=j$ &$-$   &$-$  &$-$&$\frac{(1-\epsilon)(1-p_j-q_j)}{1-q_j}$ & $\frac{\epsilon(1-p_j-q_j)}{1-q_j}$  &$-$ &$-$ &$-$ \\
${x}_i=j+1$ &$-$   &$-$  &$-$&$\frac{(1-\epsilon)p_j}{1-q_j}$ & $\frac{\epsilon p_j}{1-q_j}$  &$-$ &$-$ &$-$ \\
${x}_i=\vdots$ &$-$   &$-$  &$-$& $-$  & $-$ &$\ddots$   &$-$  &$-$ \\
${x}_i=k-1$ & $-$ & $-$ & $\cdots$ &$-$ &$-$ &$-$ &$\epsilon$ &$1-\epsilon$ \\
${x}_i=k$ & $-$ & $-$ & $\cdots$ &$-$ &$-$ &$-$ &$1-\epsilon$ &$\epsilon$ \\
\hline
\end{tabular}
\end{center}\label{tab:condprob1}
\end{table*}

\begin{prop} \label{prop:1}
For the stock-market problem described above,
\ben \begin{split} R_{ff}(D)=
\sum_{i=1}^{k-1}&\pi_i\left(h(p_i,q_i,1-p_i-q_i)-h(\epsilon,1-\epsilon)
\right)\\
& \;+\pi_k\left(h(q_k,1-q_k)-h(\epsilon,1-\epsilon)\right),
\end{split}\een
where $h()$ is the entropy function, $\left[\pi_0, \pi_1,\cdots,
\pi_k \right]$ is the stationary distribution of the Markov chain
and $\epsilon=\frac{D}{1-\pi_0}$.
\end{prop}
\begin{table*}[t]
\caption{The conditional distribution $P(\hat{X}_i|x_{i-1},x_i)$}
 \label{tab:hatx1}
\begin{center}
\begin{tabular}{|c|c|c|c|c|c|c|c|c|c|}
\multicolumn{10}{c}{$ \qquad(x_{i-1},x_i)$}\\
\hline
&$0,0$&$0,1$&$\cdots$&$j,j-1$&$j,j$&$j,j+1$&$\cdots$ &$k,k-1$&$k,k$\\
\hline
$\hat{x}_i=0$ &$1$&$1$&  $\cdots$
&$\frac{\epsilon(1-q_j-\epsilon)}{q_j(1-2\epsilon)}$
&$\frac{(1-\epsilon)(1-q_j-\epsilon)}{(1-q_j)(1-2\epsilon)}$
&$\frac{(1-\epsilon)(1-q_j-\epsilon)}{(1-q_j)(1-2\epsilon)}$&$\cdots$
&$\frac{\epsilon(1-q_j-\epsilon)}{q_j(1-2\epsilon)}$
&$\frac{(1-\epsilon)(1-q_j-\epsilon)}{(1-q_j)(1-2\epsilon)}$\\
$\hat{x}_i=1$ &$0$&$0$&  $\cdots$
&$\frac{(1-\epsilon)(q_j-\epsilon)}{q_j(1-2\epsilon)}$
&$\frac{\epsilon(q_j-\epsilon)}{(1-q_j)(1-2\epsilon)}$
&$\frac{\epsilon(q_j-\epsilon)}{(1-q_j)(1-2\epsilon)}$ &$\cdots$
&$\frac{(1-\epsilon)(q_j-\epsilon)}{q_j(1-2\epsilon)}$
&$\frac{\epsilon(q_j-\epsilon)}{(1-q_j)(1-2\epsilon)}$\\
\hline
\end{tabular}
\end{center}
\end{table*}
\begin{proof}
We will use Corollary \ref{corr:a} to verify that a first-order
Markov conditional distribution of the form \be
P_{\hat{X}_n|\hat{X}^{n-1},X^{n}}=P_{\hat{X}_n|X_n,X_{n-1}}, \quad
\forall n\ee achieves the optimum.

Due to the structure of the distortion function in Table
\ref{tab:dist1}, we choose the structure of
$P(x_i|\hat{x}_i,x_{i-1})$ as follows. When $X_{i-1}=0$, the decoder
can always declare $\hat{X}_{i}=0$ - there is no error irrespective
of the value of $X_i$. So we assign
$P(\hat{X}_i=0|x_{i-1}=0,x_i=0)=P(\hat{X}_i=0|x_{i-1}=0,x_i=1)=1$,
which gives %
%
 $P(X_i=0|x_{i-1}=0,\hat{x}_i=0)=1-p. $
The event $(X_{i-1}=0,\hat{X}_i=1)$ has zero probability. When
$(X_{i-1}=j,\hat{X}_i=0) ,\:1\leq j\leq k$, an error occurs when
$X_i=j-1$. This is assigned a probability $\epsilon$. The remaining
probability $1-\epsilon$ is split between
$P(X_i=j|x_{i-1}=j,\hat{x}_i=0)$ and
$P(X_i=j+1|x_{i-1}=j,\hat{x}_i=0)$ according to their transition
probabilities. In a similar fashion, we obtain all the columns in
Table \ref{tab:condprob1}.

We can show that the distortion criterion \eqref{eq:bindist1} can be
cast in the form
\be \label{eq:binm1} d_n(x^n,\hat{x}^n)= \frac{1}{n}\sum_{i=1}^n
\Big( -c\log_2 P(x_i|\hat{x}_i,x_{i-1})+d_0(x_{i-1},x_{i})\Big), \ee
or equivalently
\be \label{eq:binarycond1}
e(\hat{x}_i,x_{i-1},x_i)=-c\log_2
P(x_i|\hat{x}_i,x_{i-1})+d_0(x_{i-1},x_i),
\ee thereby proving that the distribution in Table
\ref{tab:condprob1} is optimal. This is done by determining the
values of $c, d_0(x_{i-1},x_{i}), \:1\leq x_{i-1},x_{i} \leq k$.
Using the values from Tables \ref{tab:dist1} and \ref{tab:condprob1}
in \eqref{eq:binarycond1}, we can find $c, d_0(.,.)$.

Since the process $\{\mathbf{X,\hat{X}}\}$ is jointly stationary and
ergodic,
%
the distortion constraint is equivalent to
$E[e(\hat{x}_2,x_{1},x_2)] \leq D.$ To calculate the expected
distortion
\be \label{eq:zz1}E[e(\hat{x}_2,x_{1},x_2)]=
\sum_{x_1,x_2,\hat{x}2}P(x_1,x_2)P(\hat{x}_2|x_1,x_2)\cdot
e(\hat{x}_2,x_{1},x_2), \ee
we need the (optimum achieving) conditional distribution
$P(\hat{X}_2|x_1,x_2)$. This is found by substituting the values
from Table \ref{tab:condprob1} in the relation
\be
P(x_2|x_1,\hat{x}_2)=\frac{P(x_2|x_1)P(\hat{x}_2|x_2,x_1)}{\sum_{x_2}P(x_2|x_1)P(\hat{x}_2|x_2,x_1)}.
\ee Thus we obtain the conditional distribution
$P(\hat{X}_2|x_1,x_2)$ shown in Table \ref{tab:hatx1}. Using this in
\eqref{eq:zz1}, we get \be E[e(\hat{x}_2,x_{1},x_2)]=
(1-\pi_0)\epsilon \leq D \ee
 We can now calculate the rate distortion function as
\be
\begin{split}
R_{ff}(D)&= \frac{1}{N}I(\hat{X}^N \to X^N)\\
&=\sum_{x_1,x_2,\hat{x}_2}P(x_1,x_2,\hat{x}_2)\log_2 \frac
{P(x_2|x_1,\hat{x}_2)}{P(x_2|x_1)}
\end{split}\ee
to obtain the expression in Proposition \ref{prop:1}.
\end{proof}

\subsection{Gauss-Markov Source}
Consider a stationary, ergodic, first-order Gauss-Markov source $X$
with mean $0$, correlation $\rho$ and variance $\sigma^2$:
\be
\label{eq:gmsource}
X_n =\rho X_{n-1} + N_n, \quad \forall n ,
\ee
where $\{N_n\}$ are independent, identically distributed Gaussian
random variables with mean $0$ and variance $(1- \rho^2)\sigma^2$.
Suppose that the source has feed-forward with delay $1$ and we want
to reconstruct at every time instant $n$ the linear combination $a
X_n +bX_{n-1}$, for any constants $a,b$. We use the mean-squared
error distortion criterion:
\be \label{eq:gmdist}
d_n(x^n,\hat{x}^n)=\frac{1}{n}\sum_{i=1}^n
\left(\hat{x}_i-(ax_i+bx_{i-1})\right)^2.
\ee
 The feed-forward distortion-rate function for this source with average mean-squared error distortion was given in \cite{tsa}.
The feed-forward rate-distortion function can also be obtained using
Theorem \ref{thm:scff} as (proof omitted)
\be R_{ff}(D)=\frac{1}{2}\log \frac{\sigma^2(1-\rho^2)}{D/a^2}. \ee
We must mention here that the rate-distortion function in the first
example cannot be computed using the techniques in \cite{tsa}.

\section{Channel Coding with Feedback}
In this section, we consider channels with feedback and the problem
of evaluating their capacity. A channel is defined as a sequence of
probability distributions: \be \label{eq:channel}
P^{ch}_{\mathbf{Y|X}}=\{P^{ch}_{Y_n|X^n,Y^{n-1}}\}_{n=1}^{\infty}.\ee
In the above, $X_n$ and $Y_n$ are the channel input and output
symbols at time $n$, respectively.
 The channel is assumed to have $k-$delay feedback
$(1\leq k < \infty)$. This means at time instant $n$, the encoder
has perfect knowledge of the channel outputs until time $n-k$ to
produce the input $x_n$. The input distribution to the channel is
denoted by
${P}^k_{\mathbf{X|Y}}=\{{P}_{X_n|X^{n-1},Y^{n-k}}\}_{n=1}^\infty.$
In the sequel, we will need the following product quantities
corresponding to the channel and the input.
 \be\label{eq:prod_dist}
\begin{split}
\vec{P}^{ch}_{Y^n|X^n}\triangleq \prod_{i=1}^n
P_{Y_i|X^{i},Y^{i-1}}, \quad
\vec{P}^k_{X^n|Y^n}\triangleq \prod_{i=1}^n
P_{X_i|X^{i-1},Y^{i-k}}.
\end{split}
\ee
The joint distribution of the system is given by
 $P_{\mathbf{X,Y}}=\{P_{X^n,Y^n}\}_{n=1}^\infty,$
 where
 $P_{X^n,Y^n}= \vec{P}^k_{X^n|Y^n}\cdot
 \vec{P}^{ch}_{Y^n|X^n}. $

\begin{defi}
An $(N,2^{NR})$ channel code with delay $k$ feed-forward  of block
length $N$ and rate R consists of a sequence of  encoder mappings
$e_i,i=1,\ldots,N$ and a decoder $g$, where \ben
\begin{split} e_i:
\{1,\ldots,2^{NR}\} \times \mathcal{Y}^{i-k} & \to
{\mathcal{X}},\quad i=1,\ldots,N\\
g:\mathcal{Y}^N & \to \{1,\ldots,2^{NR}\}\end{split} \een
\end{defi}
Thus it is desired to transmit one of $2^{NR}$ messages over the
channel in $N$ units of time. There is an associated cost function
for using the channel given by $c_N(X^N,Y^N)$. For example, this
could be the average power of the input symbols. Note that in
general, we have allowed the cost function at time $N$ to depend on
the inputs and the outputs until time $N$. This is because the
encoder knows the outputs (with some delay) due to the feedback, and
can potentially use this information to choose future input symbols
to satisfy the cost constraint.

If $W$ is the message that was transmitted, then the probability of
error is $ P_e=Pr(g(Y^N)\neq W).$
%
\begin{defi}
 $R$ is an $(\epsilon,\delta)$-achievable rate at 
 cost $C$ if for
all sufficiently large $N$, there exists an $(N, 2^{NR})$ channel
code such that
\ben P_e<\epsilon \quad \text{ and } \quad Pr(c_N(X^N,Y^N)>C)<
\delta. \een
$R$ is an achievable rate at 
cost $C$ if it is
$(\epsilon,\delta)$-achievable for every $\epsilon,\delta>0$.
\end{defi}

\begin{thm}
\emph{\cite{tati06}} \label{thm:chdir}
For an arbitrary channel $P^{ch}_{\mathbf{Y|X}}$ , the capacity with
$k-$delay feedback, the infimum of all achievable rates at
cost $C$, is given by \footnote{The $\liminf_{in prob}$ of a random
sequence $A_n$ is defined as the largest number $\alpha$ such that
$\lim_{n \to \infty}P(A_n < \alpha)=0$ and is denoted
$\underline{A}.$}
\begin{equation}\label{eq:dirch1}
C_{fb}(C)=\sup_{{P}^k_{\mathbf{X|Y}}:\rho({P}^k_{\mathbf{X|Y}}) \leq
C} \underline{I}(X \to Y),
\end{equation}
where
\ben
 {\small{ \underline{I}(X \to Y) \triangleq
\liminf_{in prob} \frac{1}{n}\log
\frac{\vec{P}^{ch}_{Y^n|{X}^n}}{P_{Y^n}} }}
\een and
{\small{
\begin{equation*}
\begin{split}
&\rho({P}^k_{\mathbf{X|Y}}) \triangleq \limsup_{in prob}
c_n(X^n,Y^n)\\
&=\inf \{ h: \lim_{n \to \infty} P_{X^nY^n}
\left((x^n,y^n):c_n(x^n,y^n)>h \right)\}=0.
\end{split}
\end{equation*}}}
\end{thm}
\vspace{3pt} In the above, we note that \ben
\begin{split}
P_{Y^n}&=\sum_{X^n}P_{X^n,Y^n}
=\sum_{X^n}\vec{P}^k_{X^n|Y^n}\cdot\vec{P}^{ch}_{Y^n|{X}^n}.
\end{split}
\een

\subsection{Evaluating the Channel Capacity with Feedback}
The capacity formula in Theorem \ref{thm:chdir} is a multi-letter
expression involving optimizing the function $\underline{I}(X \to
Y)$ over an infinite dimensional space of input distributions
${P}^k_{\mathbf{X|Y}}$.
Just like we did with sources, we can pose the following question:
\emph{Given a channel $P^{ch}_{\mathbf{Y|X}}$ and an input
distribution $P^k_{\mathbf{X|Y}}$, for what sequence of cost
measures does $P^k_{\mathbf{X|Y}}$ achieve the supremum in the
capacity formula ?}

The following theorem (proof omitted) provides an answer.
\begin{thm}
\label{thm:ccfb} Suppose we are given a channel
$P^{ch}_{\mathbf{Y|X}}$ with $k-$delay feedback and an input
distribution ${P}^k_{\mathbf{X|Y}}$  such that the joint process
$P_{\mathbf{X,Y}}$ 
is stationary, ergodic. Then the input distribution
$P^k_{\mathbf{X|Y}}$ achieves the $k-$delay feedback capacity of the
channel if for all sufficiently large $n$, the cost measure
satisfies
\be \label{eq:cost_cond} c_n(x^n,y^n)=\lambda \cdot \frac{1}{n}\log
\frac{\vec{P}^{ch}_{Y^n|X^n}(y^n|x^n)}{P_{Y^n}(y^n)}+d_0, \ee where
$\lambda$ is any positive number and $d_0$ is an arbitrary constant.
The cost constraint in this case is equal to $\limsup_{n\to
\infty}c_n(x^n,y^n)$.
\end{thm}
\appendices


\begin{thebibliography}{13}

\bibitem{shannon}
C.~E. Shannon, ``The zero-error capacity of a noisy channel,'' {\em
IRE
  Trans. Inf. Theory}, vol.~IT-2, pp.~8--19, 1956.

\bibitem{massey}
J.~Massey, ``{Causality, Feedback and Directed Information},'' {\em
Proc. 1990 Symp. on Inf. Theory and Applications (ISITA-90)},
  pp.~303--305, 1990.

\bibitem{kramer}
G.~Kramer, {\em {Directed Information for channels with Feedback}}.
\newblock Ph. D thesis, Swiss Federal Institute of Technology, Zurich, 1998.

\bibitem{sekar1}
S.~Tatikonda, {\em {Control Under Communications Constraints}}.
\newblock Ph.D thesis, Massachusetts Inst. of Technology, Cambridge, MA,
  September 2000.

\bibitem{tati06}
S.~Tatikonda and S.~Mitter, ``The capacity of channels with
feedback,'' {\em
  Submitted to IEEE Trans. Info Theory}, arXiv.org:cs.IT/0609139, 2006.

\bibitem{tsa}
T.~Weissman and N.~Merhav, ``On competitive prediction and its
relation to
  rate-distortion theory,'' {\em IEEE Trans. Inf. Theory},
  vol.~IT-49, pp.~3185--3194, December 2003.


\bibitem{pradhan07}
S.~S.Pradhan, ``On the Role of Feedforward in Gaussian Sources:
Point-to-Point Source Coding and Multiple Description Source
Coding,'' {\em IEEE Trans. Inf. Theory}, vol.~53, no.~1,
  pp.~331--349, January 2007.

\bibitem{mw}
E.~Martinian and G.~W. Wornell, ``{Source Coding with Fixed Lag Side
  Information},'' {\em Proc. 42nd Annual Allerton Conf.
  (Monticello, IL)}, 2004.

\bibitem{rdff}
R.~Venkataramanan and S.~S. Pradhan, ``{Source coding with
feed-forward:
  Rate-distortion theorems and error exponents for a general source},'' {\em
  Proc. IEEE Inf. Theory Workshop, San Antonio, 2004; To appear in IEEE
  Trans. Inf. Theory, 2007}.

\bibitem{hv}
T.~Han and S.~Verd\'{u}, ``Approximation theory of output
statistics,'' {\em
  IEEE Trans. Inf. Theory}, vol.~39, pp.~752--772, May 1993.

\bibitem{ck}
I.~Csisz'ar and J.~Korner, {\em Information Theory: Coding Theorems
for
  Discrete Memoryless Systems}.
\newblock New York: Academic Press, 1981.

\bibitem{gast}
M.~Gastpar, B.~Rimoldi, and M.~Vetterli, ``To code, or not to code:
lossy
  source-channel communication revisited.,'' {\em IEEE Trans. Inf. Theory}, vol.~49, no.~5, pp.~1147--1158, 2003.

\bibitem{duality}
S.~S. Pradhan, J.~Chou, and K.~Ramchandran, ``Duality between source
coding and
  channel coding and its extension to the side-information case,'' {\em IEEE
  Trans. Inf. Theory}, vol.~49, pp.~1181--1203, May 2003.

%

\end{thebibliography}
\end{document}